\documentclass{aa}
\usepackage{graphicx}
\usepackage{txfonts}
\begin{document}
\title{The spectral-type/luminosity and the 
spectral-type/satellite-density relations in the 2dFGRS}
\author{B. Kelm \inst 1, P. Focardi \inst 1 and G. Sorrentino \inst 2} 
\institute{Dipartimento di Astronomia, Universit\`a di Bologna, V. Ranzani 1, 
I-40127 Bologna
\and
INAF-Osservatorio Astronomico di Capodimonte, V. Moiariello  16, I-80131 Napoli} 
\abstract{We examine the relative fractions of passive 
(Type 1), quiet-SF (Type 2) and active-SF (Type 3+4) galaxies as a function of 
luminosity and number of neighbours in several  volume-limited samples selected from the 2dFGRS. 
Neighbours are counted within 1 $h_{75}^{-1}$ \,Mpc projected distance 
and $\pm$ 1000 km\,s$^{-1}$ depth. We apply a maximum magnitude difference 
criterion and require neighbours to be fainter than the galaxy itself.
 We show that, whatever the environment, passive galaxies dominate in bright samples  and active-SF galaxies in faint samples, whereas quiet-SF galaxies 
never dominate. We further show that in bright samples (M$_{B}$\,--\,5 $\log$ $h_{75}$ $\leq$$ \,-\,19$) the fraction of passive galaxies grows steadily 
with fainter neighbour density, whereas in faint samples a threshold-like 
dependence is observed. This suggests that the spectral-type / density 
($\approx$ morphology / density) relation extends to the intermediate dense 
environment, but only in the surroundings of luminous galaxies and that it 
reflects an enhancement of the number of satellites rather than stronger 
clustering among galaxies themselves. Our analysis indicates that, in general,
 luminosity is a good tracer of galaxy halo mass and  that it dominates over 
environment (satellite density) in setting the spectral type 
mix of a population. However, minority populations exist, such as luminous 
SF galaxies and faint passive galaxies, whose luminosity is an inaccurate  
tracer of halo mass. 
}
\titlerunning{Type-density-luminosity in the 2dFGRS}
\maketitle
\section{Introduction}
The existence of a relation between galaxy type and environment dates back 
to pioneering work by Hubble\& Humason (1931) 
and was first quantified by Davis \& Geller (1976) and Dressler (1980). 
Dressler found that, in nearby clusters, the fraction of elliptical galaxies 
increases and the fraction of spirals decreases  with increasing 
density. 
Whitmore \& Gilmore (1991) and Whitmore et al. (1993) suggested that the 
correlation between morphology 
and cluster-centric-radius is tighter than the correlation between 
morphology and density. 
Deep surveys such as the 2dFGRS and the SDSS have confirmed 
 that the star-formation level in galaxies decreases at large galaxy 
density (\cite{Lewis02,Gomez03,Goto02,Goto03,Balogh04a,Christlein05}) and 
that a threshold is reached at low densities 
($\approx$ 1 gal Mpc$^{-2}$, M$_{B}$ $\leq$ --19), below which no further increase in 
star-formation is observed. 
These results provide evidence that, at least at low redshift, 
a strong correlation exists between the characteristics of the stellar 
component of a galaxy and its surrounding environment, with 
luminosity and color being the galaxy properties 
most strongly correlated with environment (\cite{Kauffmann04,Blanton05a}). 
But the exact dependence of the morphology/density relation on density and
on luminosity 
is still a matter of debate.  
It is not clear yet whether the relation 
extends to galaxies in systems less dense than clusters 
(i.e. to the majority of galaxies). 
Postman \& Geller (1984) and Maia \& da Costa (1990) claim that the 
morphology/density relation certainly extends to groups, 
whereas Whitmore (1995) does not confirm their result. 
Tran et al. (2001) and Helsdon \& Ponman (2003) 
support the existence of the morphology/density relation in X-ray bright 
groups. Dominguez et al. (2002) show that the 
relation is observed in very massive (optically selected) groups only, 
whereas Kelm \& Focardi (2004a) report that 
the frequency of early-type galaxies is larger in Compact Groups 
than among isolated galaxies. Mateus \& Sodr\'e (2004) and Gerken et al. (2004) 
provide evidence that, even outside clusters, star formation properties 
are affected in all ranges of density. Tanaka et al. (2004) find
that only faint galaxies show a break in star formation and morphology at a 
critical local density.   
 Also the role of luminosity on the morphology/density relation 
 has still to be  fully disentangled. 
Galaxy clustering appears to depend on luminosity for luminous galaxies 
and on color for low luminous ones (\cite{Norberg02a,Hogg03,Balogh04b,Berlind04,Zehavi04}). For blue galaxies the relation between environment and 
luminosity is typically weak, whereas for red galaxies clustering is likely 
a non-monotonic function of luminosity, peaking at both high and low 
luminosities.  

In this paper we investigate the relation linking the galaxy spectral type mix 
with both luminosity and local environment. 
Specifically we explore 1) how the spectral-type/luminosity 
relation varies as a function of environment and 2) how the 
spectral-type/satellite-density relation varies as a function of 
luminosity.  We use data from the 2dF 
to select 10 different volume limited samples, covering a wide luminosity range 
(--22.5 $\leq$ M$_{B}$\,-5 $\log$ $h_{75}$  $\leq$ --17.0).
We evaluate for each galaxy the neighbour density on the characteristic scale of galaxy 
groups ($\sim$\,1\, $h_{75}^{-1}$ Mpc) which further corresponds to the present day 
typical virial radius of halos. 
Galaxy properties are expected to correlate most strongly with densities 
evaluated on this scale, also from a theoretical standpoint 
(\cite{Blanton05b,Kauffmann04,Berlind04}).  

At variance with previous analysis we compute neighbour density applying a 
maximum magnitude difference criterion and count neighbours over a 2 
magnitude interval.   
The adopted range in magnitude reduces the number of galaxies that have no 
neighbours, on a group scale, to $\simeq$15\% and associates  
most galaxies (2/3) with a number of neighbours (1$\leq$neigh$\leq$8) that  
matches the typical observed environment of galaxies in groups. 
We also limit neighbour computation to galaxies that 
are fainter than (or equally luminous to) the galaxy itself. Usually, 
when computing density in volume-limited samples, no distinction is made 
between brighter and fainter neighbours. 
This implies that, within the same volume-limited sample, the density definition depends on luminosity: 
the environment of luminous galaxies is defined by fainter neighbours 
whereas the environment of low luminous galaxies is defined
by brighter neighbours. 
But, obviously, the impact of a brighter or a fainter companion  
on a galaxy is different.
Less massive companions have likely been, or will be, accreted by the galaxy halo, 
whereas more massive companions will likely accrete the galaxy and destroy its halo.

Our density definition is luminosity-independent. 
For luminous
galaxies, the environment on the 1 h$_{75}^{-1}$ Mpc scale is likely to 
correspond to the density 
of satellites that have been captured by the galaxy  halo. Conversely, for
low luminous galaxies, it likely corresponds to the clustering of small
halos among themselves, or, in the case of a galaxy swallowed up in the halo of a bigger companion, 
for the richness of satellites within this large halo.
 
In $\S$2 we present the sample, in $\S$3 we discuss the link 
between density distribution and luminosity. 
In $\S$4 and $\S$5 we explore the dependence on density of the 
spectral-type/luminosity relation. 
In  $\S$6 we investigate the dependence on luminosity of the 
spectral-type/density relation. 
The summary and conclusions are given in $\S$7. 
We assume $\Omega_M=0.3$, $\Omega_{\Lambda}=0.7$, and 
$h_{75}$ = H$_o$ /(75 km\,s$^{-1}$\,Mpc$^{-1}$) = 1. 
\section{The sample}
The sample we use for the present analysis is  
selected from the 2dFGRS (\cite{Colless01,Colless03}). 
The 2dF covers $\sim$ 1800 square degrees and is complete for 
galaxies down to an extinction-corrected limit of $b_J$ = 19.45. 
It provides 
redshifts, in the range 0 $\leq$ $z$ $\leq$ 0.3, for 221,496 galaxies selected 
from the APM catalogue (\cite{Norberg02b}),  
which is 90-95\% complete (\cite{Maddox90}).  
Because saturation effects and stellar contamination cannot be ignored 
for bright galaxies, we exclude from the sample galaxies brighter  
than b$_j$ = 16.

Each 2dF galaxy spectrum is typed on the basis of the relative 
strength of its first two principal components 
(for details on the PCA see Folkes et al. 1999), which are the emission 
and absorption components within the spectrum. 
The parameter $\eta$ (\cite{Madgwick02})  
is the linear combination of these two components.   
Qualitatively $\eta$ is an indicator of the ratio of the present to the past 
star-formation activity of each galaxy, but it is  
reliable only for z $\leq$ 0.15. 
Clusters are dominated by galaxies with the lowest $\eta$ values, 
whereas the field contains a much larger proportion of galaxies with higher 
($\eta > $0) values. 
The median $\eta$ correlates with morphological classes, (low $\eta$ are 
typically early type galaxies, high $\eta$ late type galaxies) 
although there is a large scatter in the $\eta$ values of spectra that lie 
within a given morphological class. 

As in Madgwick et al. (2002) we divide the $\eta$ scale into 4 intervals: 
\[
 \eta <  -1.4\, \Longrightarrow \, Type\, 1
\]
\[
 -1.4 \leq \eta < 1.1\, \Longrightarrow \, Type\, 2
\]
\[ 
 1.1 \leq\eta < 3.5\, \Longrightarrow \, Type\, 3
\]
\[
 \eta \geq 3.5\, \Longrightarrow \, Type\, 4
\]

We group together Type 3 and Type 4 galaxies, the latter being rare.
Throughout the paper, Type 1, Type 2 and Type 3\,+\,Type 4 
galaxies are named passive, quiet-SF and active-SF respectively.  
We keep quiet-SF and active-SF galaxies separate, in order to investigate 
any dependence of SF triggering processes on specific density 
characteristics.  

For each 2dF galaxy (random fields excluded) with  z $\leq$ 0.15 
we have automatically identified neighbours 
within 1\,$h_{75}^{-1}$ Mpc projected distance and  $\pm$ 1000 km\, s$^{-1}$ depth. 
We count as neighbours all galaxies fainter than the galaxy itself 
that satisfy a maximum magnitude difference criterion 
(--2 $\leq$ $M_{gal} - M_{neigh}$ $\leq$0). 
We reject from the sample all galaxies 
whose 2 magnitude fainter companions would fail the 2dF selection 
criteria. 
The minimum fiber separation of the  2dF survey ($\sim$ 30 \arcsec) tends to 
reduce the number of close neighbours of galaxies, a bias that might affect 
passive galaxies more severely than SF galaxies.  
However, the bias is likely marginal as this separation corresponds to less 
than one-tenth of the explored distance even for the highest redshift galaxies.

The final sample includes $\sim$14,000 galaxies in the redshift 
range 0.0156 $\leq$\,z\,$\leq$ 0.15, and  absolute magnitude 
range --22.5 $\leq$ M$_{B}$\,-5 $\log$ $h_{75}$  $\leq$ --17.0.
Absolute magnitudes are computed adopting the 
k-correction as in Magdwick et al. (2002), which varies with  
galaxy spectral-type.      
We split the sample into 10 different volume-limited subsamples, 
covering a 1 magnitude range each and overlapping by 0.5 magnitude. 
Since the k-corrections are class 
dependent, the z$_{min}$ and z$_{max}$ 
values corresponding to a given absolute magnitude range are also class 
dependent. Hence, the volumes defining the samples for two different 
spectral classes, for the same bin in absolute magnitude, will not exactly 
coincide (see also Norberg et al. 2002a.).  

Table 1 lists the spectral-type composition in each  
volume limited sample.   
The gap between galaxies in the faintest and the brightest volume limited 
samples corresponds to a factor $\sim$150 in luminosity.    
\begin{table}
\begin{center}
\caption[] {Spectral-type content of the volume limited samples. 
Each sample spans a one magnitude range, and overlaps by 0.5 magnitude with the next one. 
M$_B$ corresponds to the central value of the magnitude in each bin. }
\begin{tabular}{||r||r|r|r||r||}
\noalign{\smallskip}
\hline
\hline
\noalign{\smallskip}
M$_B$ & N$_{T1}$ & N$_{T2}$ & N$_{T3+T4}$ & N$_{tot}$\cr 
\hline
\hline
\noalign{\smallskip}
$-22.0$     &  782  &  249  &   41 & 1072 \\ 

$-21.5$    & 1865  &  699  &  144 & 2708 \\ 

$-21.0$     & 2703  & 1395  &  421 & 4519 \\ 

$-20.5$     & 2732  & 1908  &  828 & 5468 \\ 

$-20.0$     & 2134  & 1860  & 1100 & 5094 \\ 

$-19.5$     & 1194  & 1183  &  918 & 3295 \\ 

$-19.0$     &  543  &  692  &  773 & 2008 \\ 

$-18.5$     &  277  &  467  &  663 & 1407 \\ 

$-18.0$     &  158  &  327  &  451 & 936 \\

$-17.5$     &   77  &  189  &  276 & 542 \cr
\hline
\hline
\end{tabular}
\end{center}
\end{table}
\section{The dependence on luminosity of the satellite-density distribution}
We assign a local density (number of fainter neighbours within 
1 $h_{75}^{-1}$ Mpc, $\pm$ 1000 km$^{-1}$ depth and a 2-$\Delta$Mag range) 
to all galaxies in our sample. 
We then define four distinct density regimes characterized by different number of neighbours: 

\[
 neigh = 0
\]
\[
 1 \leq neigh \leq 2 
\]
 \[
 3 \leq  neigh \leq 8 
\]
\[
 neigh >  8
\]

Table 2 lists the number of galaxies per spectral-type 
in the four defined density ranges and for each volume limited sample. 
The density parameterization is such that 
most galaxies ($\sim$ 2/3) are in the two central bins, 
which exhibit the typical galaxy density of groups. 
Obviously, computing densities on the galaxy group scale 
does not correspond to selecting a sample of 2dF groups 
(\cite{Eke04a,Eke04b,Merchan02}). 

\begin{figure}
\resizebox{\hsize}{!}{\includegraphics{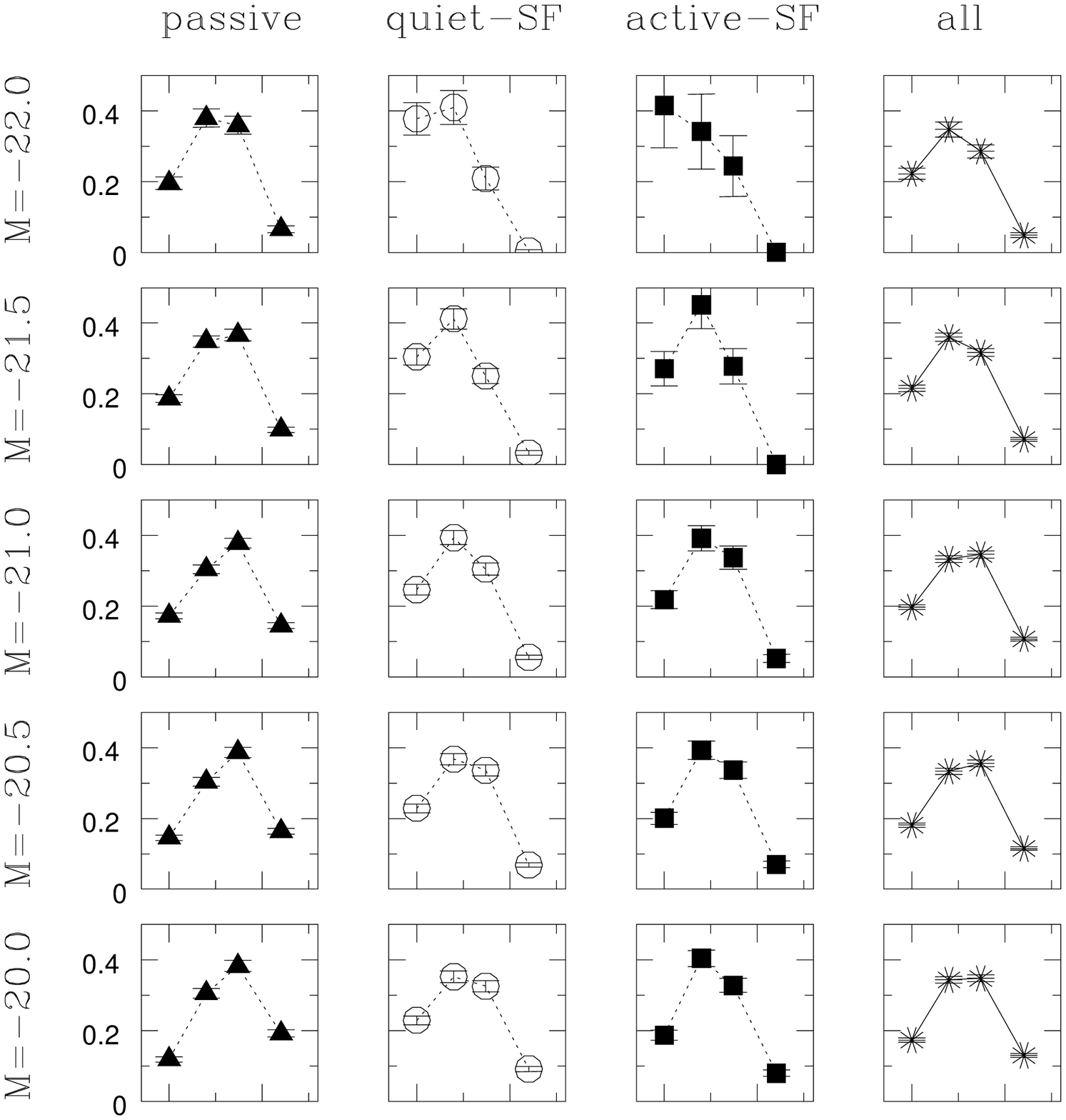}}
\resizebox{\hsize}{!}{\includegraphics{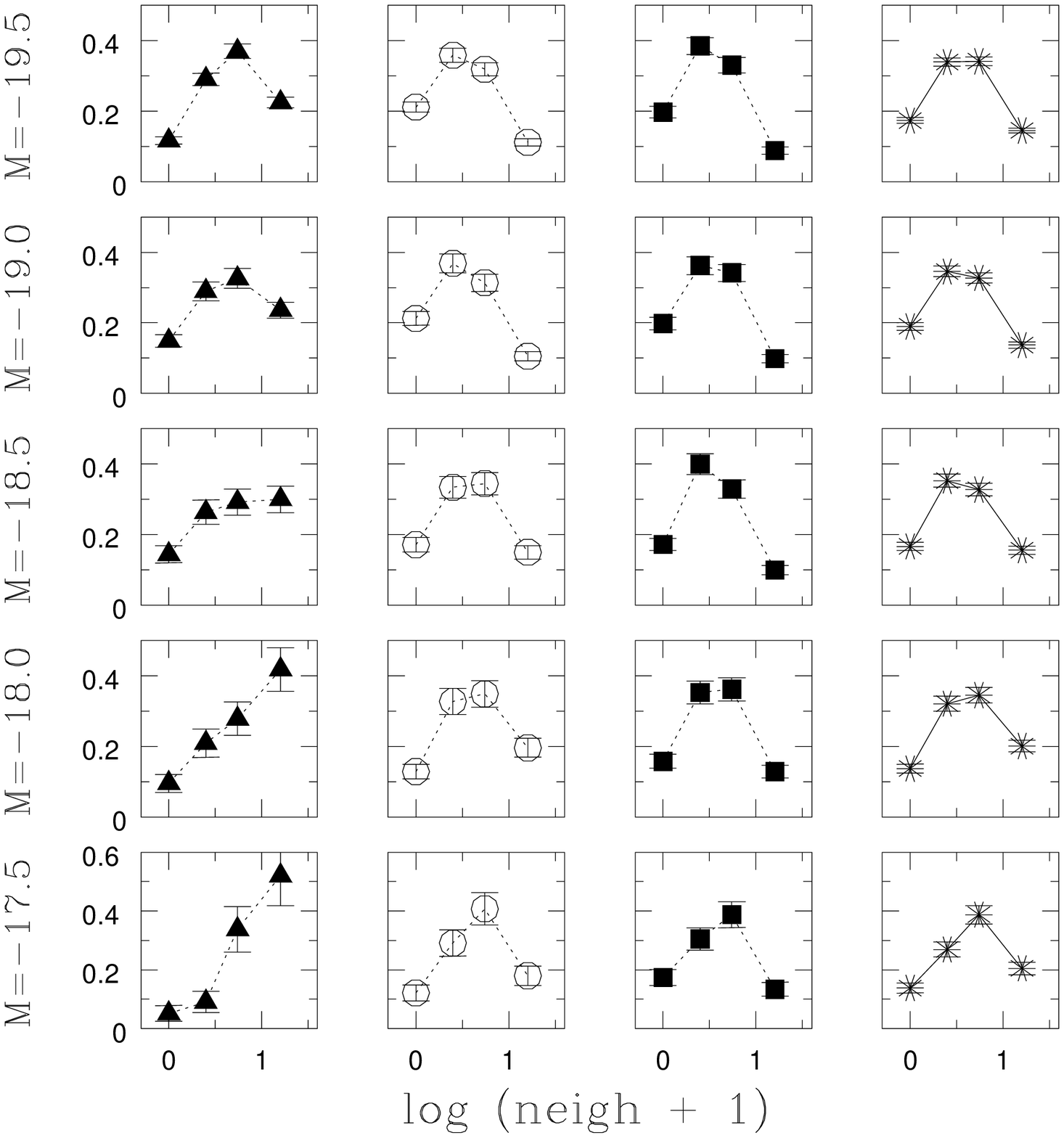}}
\hfill
\caption{Relative distributions of passive (Type 1), quiet-SF 
(Type 2), active-SF (Type 3+4) and all-Type galaxies as a function of 
number of fainter neighbours (in 4 intervals), for the 10 volume 
limited samples. Numbers are normalized to the total number of galaxies 
of a given type: for each spectral type the sum over all environments is 1. 
The number of neighbour distribution, in the all-Type galaxy 
sample, undergoes no significant modification with luminosity.
}
\end{figure} 

\begin{table*}
\begin{center}
\caption[] {Spectral-type content of the volume-limited samples as a function of neighbour density.}
\begin{tabular}{||r||rrrr||rrrr||rrrr||rrrr||}
\noalign{\smallskip}
\hline
 &\multicolumn {4 }{c|| }{neigh=0}& \multicolumn {4 }{c||}{1 $\leq$ neigh$\leq$2} & 
\multicolumn {4 }{c||}{3$\leq$ neigh $\leq$ 8} & \multicolumn {4 }{c||}{neigh$>$8}\\
\cline{2-17}
\noalign{\smallskip}
M$_B$ & T$_1$ & T$_2$ & T$_{3+4}$& $\Sigma$  & T$_1$ & T$_2$ & T$_{3+4}$ &$\Sigma$ &  T$_1$ & T$_2$ & T$_{3+4}$ & $\Sigma$  & T$_1$ & T$_2$ & T$_{3+4}$ & $\Sigma$\cr
\hline
\noalign{\smallskip}

$-22.0$  &   153  &   94 &  17 &  264    &  297  &  102 &       14 &   413   & 281  &   52 &    10   &  343     &  51   &   1    &  0 &    52   \\
$-21.5$  &  348  &  213 & 39 & 600    &  649  &  288 &      65 & 1002   &  684  &  175 &   40 &  899     & 184   &  23   & 0  &   207   \\
$-21.0$  &  466  &  344 &  92 & 902    &  823  &  549 &  165 & 1537   &  1022  &  425 &  142 & 1589     & 392   &  77   &  22&   491   \\
$-20.5$  &  398  &  436 & 166  & 1000  &  830  &  701 &  325  & 1856   & 1057  &  641 &  279 & 1977     & 447   & 130   &  58&   635   \\
$-20.0$  &  254  &  427 & 206  & 887   &  652  &  656 &  445  & 1753   &  817  &  606 &  361 & 1784     & 411   & 171   &  88&   670   \\
$-19.5$  &  139  &  250 & 181 & 570    &  346  &  424 &  353  & 1123   &  441  &  377 &  303 &  1121     & 268   & 132   &  81&   481   \\
$-19.0$  &   81  &  147 & 153 & 381    &  157  &  255 &  280  &  692   &  177  &  217 &  264 &  658     & 128   &  73   &  76&   277   \\
$-18.5$  &   40  &   80 &  114  & 234   &   73  &  156 &  265  & 494   &   81  &  161 &  218 &  460     &  83   &  70   &  66&   219   \\
$-18.0$  &   15  &   42 &  71 & 128    &   33  &  107 &     159  &   299  &   44  &  114 &  163  & 321      &  66  &  64   &  58&   188  \\
$-17.5$  &   4  &   23 &  48 & 75    &   7  &  55 &   84  &   146  &   26  &  77 &  107 & 210      &  40  &  34   &  37&   111  \cr
\hline
\noalign{\smallskip}
\end{tabular}
\end{center}
\end{table*}
The relation linking luminosity and fainter neighbour density  
is shown in Fig. 1. Distributions of    
passive, quiet-SF, active-SF and all-type galaxies,
normalized to the total number of galaxies of a given type, 
are shown, for the 10 volume-limited samples.   
The last column of Fig.1 shows that in a composite (all-type) 
population, the number of fainter neighbours associated 
with galaxies is a weak function of luminosity. 
This recalls the result by Zehavi et al. (2002), showing that 
all-type galaxy subsamples in 3 distinct absolute magnitude ranges have 
real-space correlation functions that are parallel power-laws. 

If we assume that for luminous galaxies, 
the environment on a 1 $h_{75}^{-1}$ Mpc scale essentially stands for the density 
of satellites that have been captured by the galaxy halo  
whereas for faint galaxies, 
it stands for the number of neighbour 
galaxies still in their own small halo, then Fig. 1 actually indicates that 
the distribution of satellites surrounding luminous central galaxies within 
large halos, and the distribution of fainter companions surrounding low 
luminous galaxies, are almost self similar. 
This implies that without information on the luminosity of the galaxies, 
the neighbour density distribution of galaxies (on 1 $h_{75}^{-1}$ Mpc scale) cannot
be used to discriminate between massive group-size halos and associations of galaxies 
in distinct small-size halos.   

Figure 1 also clearly shows that at all luminosities, 
passive and SF galaxies exhibit different distributions
and that the excess of companions surrounding passive 
galaxies is not limited to luminous galaxies 
(\cite{Norberg02a,Hogg03,Berlind04,Blanton05a})
but is instead a general characteristic of passive galaxies. 
The all-type galaxy distribution reflects 
the passive population at the bright end, and the star-forming galaxy 
population at the faint end. 
Our assumption that luminous galaxies are central galaxies within group-size 
halos and faint galaxies are central galaxies within small size halos is 
therefore further consistent with 
the expectation that the SFR of a galaxy is a decreasing 
function of its halo mass . 

However SF galaxies are found among bright galaxies and passive galaxies 
among faint ones.
How can we explain their existence? 
We will assume that luminous SF galaxies are hosted in small mass halos; 
they may exhibit several neighbours, but, at variance with luminous 
passive galaxies, neighbours are not embedded within the galaxy halo. 
As a consequence optically 
selected passive dominated groups are predicted to be systematically 
more massive than optically selected SF dominated 
groups (\cite{Kelm04a,Kelm04b,Mulchaey03}). 
Similarly, we explain the existence of low luminous
passive galaxies assuming that they are satellites embedded within the halo 
of a large (group-size) system. Actually, the large fraction of faint passive
 galaxies with $>$8 neighbours among faint galaxies (see Table 2)  suggests 
 that this population is tracing a 
large potential well (\cite{Norberg02a,Hogg03,Berlind04,Zehavi04,Jing04}), 
with the galaxy and all of its 
fainter neighbours having been accreted by a massive system. 
 
The stronger clustering of passive galaxies relative to SF galaxies, 
on the group scale, appears to arise from two distinct contributions. 
At the luminous end, it is due to an excess of satellites surrounding 
central galaxies inside large halos. At the faint end it is due to an excess of 
satellites that are strongly correlated among themselves.  

\begin{figure}
\resizebox{\hsize}{!}{\includegraphics{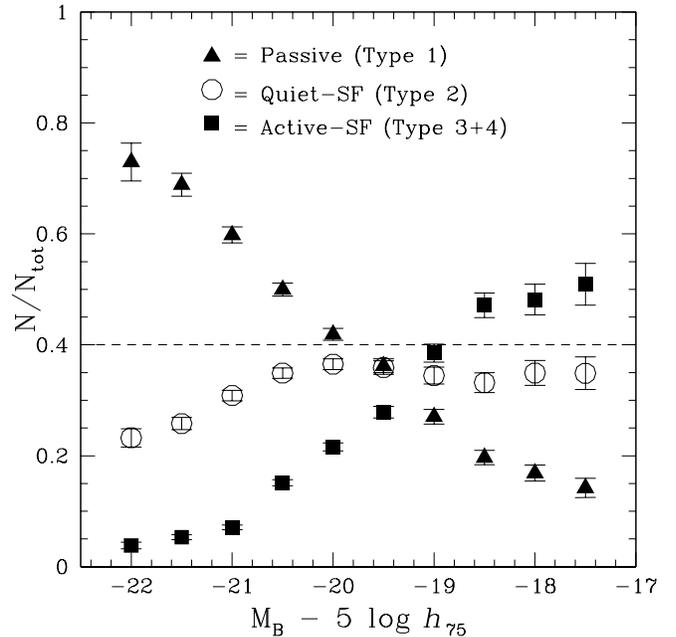}}
\hfill
\caption[]{The relative content in passive, quiet-SF and 
active-SF galaxies in the 10 volume-limited samples. 
Error bars are multinomial. 
The 40\% fraction line is drawn to reveal 'dominant' populations.    
Passive galaxies dominate in bright samples, active-SF  
galaxies in faint samples, quiet-SF galaxies are never dominant. 
The contribution of the 3 populations is comparable in the 
M$_B$ -- 5 $\log$ $h_{75}$  $\in$ [$-20 \div -19$] range.  
}
\end{figure} 
\begin{figure}
\resizebox{\hsize}{!}{\includegraphics{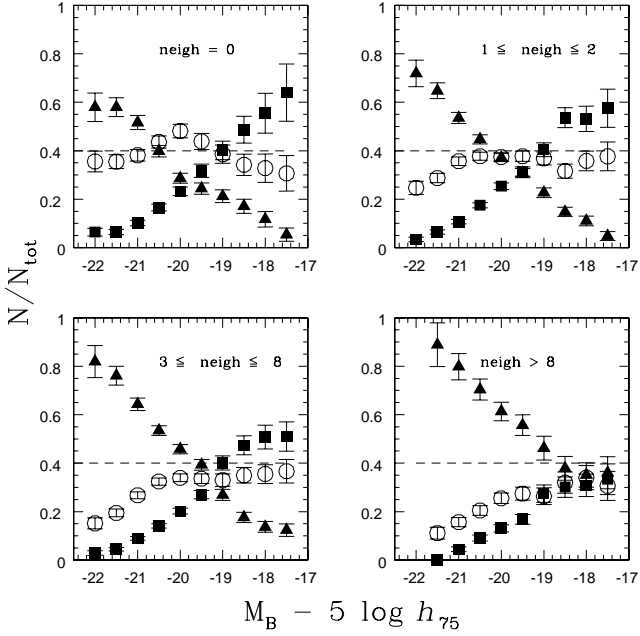}}
\hfill
\caption[]{Relative content in passive, quiet-SF and active-SF 
galaxies in the 10 volume-limited samples (symbols as in Fig.2).  
Each panel refers to galaxies in a specific neighbour density bin. 
The absence of satellites (neigh=0) efficiently reduces the 
dominance of passive galaxies in luminous samples, raising the contribution 
of quiet-SF galaxies. Conversely, a high density environment (neigh $>$8)
significantly enhances the contribution of passive galaxies in 
faint samples. 
}
\end{figure}
\section {The dependence on satellite-density of the 
spectral-type/luminosity relation} 
Figure 2 shows the fractions of passive, quiet-SF and active-SF 
galaxies in  the 10 volume-limited samples  for the 
total (all neighbour density) population. 
It is a global spectral-type/luminosity plot that indicates how 
the fraction of passive and SF galaxies varies as a function of  
luminosity. It is similar to Fig. 9 in Norberg et al. (2002a),  
in which fractions have been derived for late-type and early-type galaxies 
only.  
   
The horizontal line in Fig. 2 denotes the 40\% fraction: 
points above this threshold mark dominant populations. 
Passive galaxies are 'dominant' in galaxy samples 
brighter than M$_{B}$ - 5 $\log$ $h_{75}$ = -20, active-SF  galaxies  
are 'dominant' in samples fainter than M$_{B}$ - 5 $\log$ $h _{75}$ = --19. 
 Contributions from different spectral-type populations are comparable in 
the [$-20 \div -19$] magnitude bin.   
These trends 
confirm that star-formation activity  in the local universe 
definitely is a characteristic of low luminosity galaxies.    

In Fig. 3 we break down the contributions 
of passive and SF galaxies to the Fig. 2 plot into their contributions 
from systems exhibiting different numbers of fainter neighbours.   
This allows us to explore how strongly the relative fraction of passive and SF 
galaxies depends on environment. 
If the dominance of passive galaxies at high luminosity  and 
the dominance of active-SF galaxies at low luminosity were 
independent of neighbour density, we would expect all panels in Fig.\,3 to 
be similar. 
This is not the case, 
however, differences among panels are modest: 
whatever the number of satellites, bright samples are 
dominated by passive galaxies, and faint samples by active-SF ones. 

Therefore, in general, luminosity dominates over  neighbour 
multiplicity in setting the spectral-type mix of a galaxy population. 
A luminous galaxy might have few or many satellites, but will  
likely trace a deep potential. 
A faint galaxy might have few or many neighbours but will likely trace a shallow potential. 
It is only in extreme environments that the mix set by luminosity is significantly modified:
galaxies with  neigh$>$8 have their star-formation level (a typical active-SF one) 
suppressed even in faint samples, whereas isolated (neigh=0) galaxies are 
still 40\% likely to be star-forming (quiet-SF) 
at M$_{B}$\,--\, 5 $\log$ $h_{75}$\,$\simeq$\,$-21.5$. 
These trends are consistent with result discussed in $\S$3, namely 
that ''minority'' population can be identified in very 
luminous and very faint samples whose luminosities  
are inaccurate tracers of their halo mass: 
luminous SF galaxies trace small (sub-group size) halos, 
whereas faint passive galaxies trace massive (group/cluster size) 
halos. 

Figures 2 and 3 also provide evidence that fractions of active-SF and quiet-SF 
galaxies exhibit distinct trends with luminosity: the fraction  
of active-SF galaxies decreases towards increasing luminosity, 
while the fraction of quiet-SF galaxies is nearly independent of 
luminosity, except for the most luminous samples. 
While Fig.1 indicates that the dependence on density is the same 
for quiet-SF and active-SF galaxies (\cite{Madgwick03}), Fig. 2 and Fig. 3 
indicate that the dependence on luminosity is different. 
The data thus suggest a bimodal behaviour 
for  galaxies with satellite-density and a 'trimodal' behaviour 
(passive, quiet-SF, and active-SF) with luminosity.
Bimodality in the distribution of galaxies properties has been addressed 
in many recent papers (\cite{Strateva01,Hogg02,Balogh04b,Berlind04,Blanton05a}).
\section{The role of satelite density for passive and SF galaxies} 
To further explore the trend of increasing passive and decreasing SF 
galaxy fraction with luminosity and fainter neighbour density we 
also show, in Fig. 4, the fractional content of passive, quiet-SF and 
active-SF galaxies in different environments. 
The relative role of extremely dense and intermediate dense 
environment can be easily explored by comparing the gap between the neigh=0 
and the neigh$>$8 lines with the the gap between the neigh=0 and the 
neigh=3-8 lines. 
Figure 4 shows that, for all 3 types, the gaps undergo a strong 
variation at magnitude M$_ {B}$ - 5 $\log$ $ h_{75}$ $\approx$ --19. Therefore, we 
will keep the analysis of faint and bright samples separate. 
\begin{figure}
\resizebox{\hsize}{!}{\includegraphics{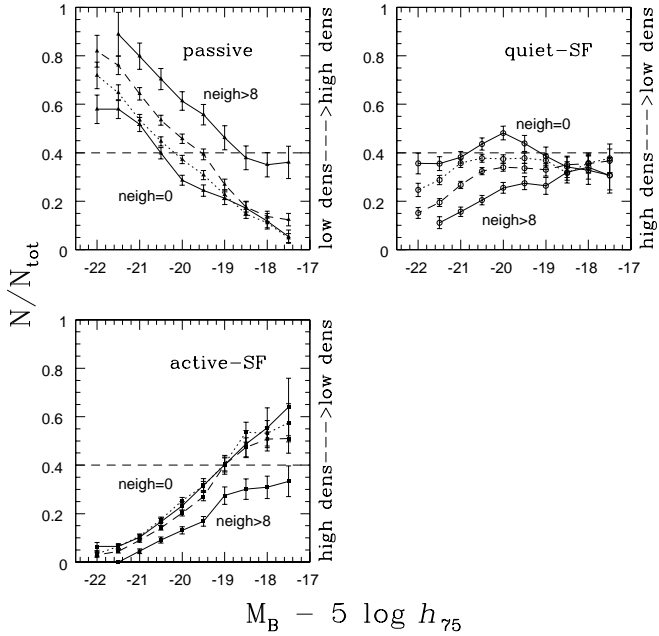}}
\hfill
\caption[]{Relative content in passive (top-left), quiet-SF (top-right) 
and active-SF (bottom-left) galaxies in samples of galaxies 
in different environments. 
Continuum lines are used to connect galaxy fractions 
with 0 and $>$8 neighbours, dotted and hatched lines to connect galaxy 
fractions in intermediate dense environments (1 $\leq$ neigh $\leq$ 2 and 3 $\leq$ neigh $\leq$ 8).  
}
\end{figure}
Figure 4 shows that in luminous samples the fraction of 
isolated passive galaxies (neigh=0) is always below the fraction of passive 
galaxies with neigh$>$8, and that the gap does not depend on luminosity.  
A similar specular large gap is observed for quiet-SF galaxies,  
whereas the gap is smaller for active-SF galaxies. 
A smaller, but still significant, gap is observed between the neigh=0 and the 
neigh=3-8 lines, in passive and quiet-SF galaxy samples. The 
size of the gap is, again, nearly luminosity independent. Conversely,   
no gap is associated with active-SF galaxies. 

In faint samples (M\,-5 $\log$ $h_{75}$$\geq$ --19), passive and active-SF 
galaxies exhibit a large gap between the neigh=0 and the neigh$>$8 
lines, whereas no gap is associated with the neigh=0 and neigh=3-8 transition. 
Again, a distinct behaviour is observed for active-SF and quiet-SF galaxies,  
the latter being equally frequent in all environments. 

In summary, Fig. 4 indicates that a continuous parameterization of 
neighbour multiplicity (from 0 to 1-2 to 3-8 to $>$8) is indeed meaningful 
for bright samples, where neighbours are mainly satellites, as it relates to different 
fractions of passive and quiet-SF galaxies. 
In faint samples, however, a threshold-like density 
parameterization appears to describe the galaxy behaviour better 
than a continuous one. 
This suggests that a continuous relation linking spectral-type with 
density only occurs when computing the density of satellites surrounding
very luminous galaxies and implies that the spectral-type/density relation
traces an enhanced correlation inside massive halos rather than an enhanced 
correlation between distinct halos.  

The M$_B$ - 5 $\log$ $h_{75}$ $\simeq-19$ magnitude is a critical one: 
 it corresponds to the luminosity at which the dependence on 
 satellite-density moves from continuous to threshold-like, and also 
to the luminosity 
where fractions of active-SF galaxies become larger than the fractions of quiet-SF 
galaxies (see Fig. 2). Therefore it corresponds to the luminosity above 
which samples of passive galaxies exhibit a dependence on 
 fainter neighbour 
density that is specular relative to that of quiet-SF galaxies, 
and below which passive galaxies are specular to active-SF galaxies. 
This is consistent with the 
finding (\cite{Norberg02a,Balogh04a}) that, for low luminosity galaxies, 
clustering is a strong function of color, while for luminous galaxies 
clustering is a strong function of luminosity. 
\section {The dependence on luminosity of the spectral-type/neighbour 
density relation}
We have shown that the spectral-type/density relation is possibly 
a spectral-type/satellite-density relation that traces an enhanced correlation 
inside single massive halos rather than enhanced correlation between 
distinct halos. To test this assumption directly 
we next examine the dependence of the spectral-type/density 
($\sim$morphology/density) relation on luminosity.     
In Fig. 5 we show the spectral-type/density relation, with fractions 
of passive, quiet-SF, active-SF and all-SF galaxies normalized to the total 
number of galaxies in a given density bin, for  9 volume-limited samples. 
The faintest sample is not shown because it is small (see Table 2) 
and more affected by statistical uncertainties.  
\begin{figure*}
\resizebox{\hsize}{!}{\includegraphics{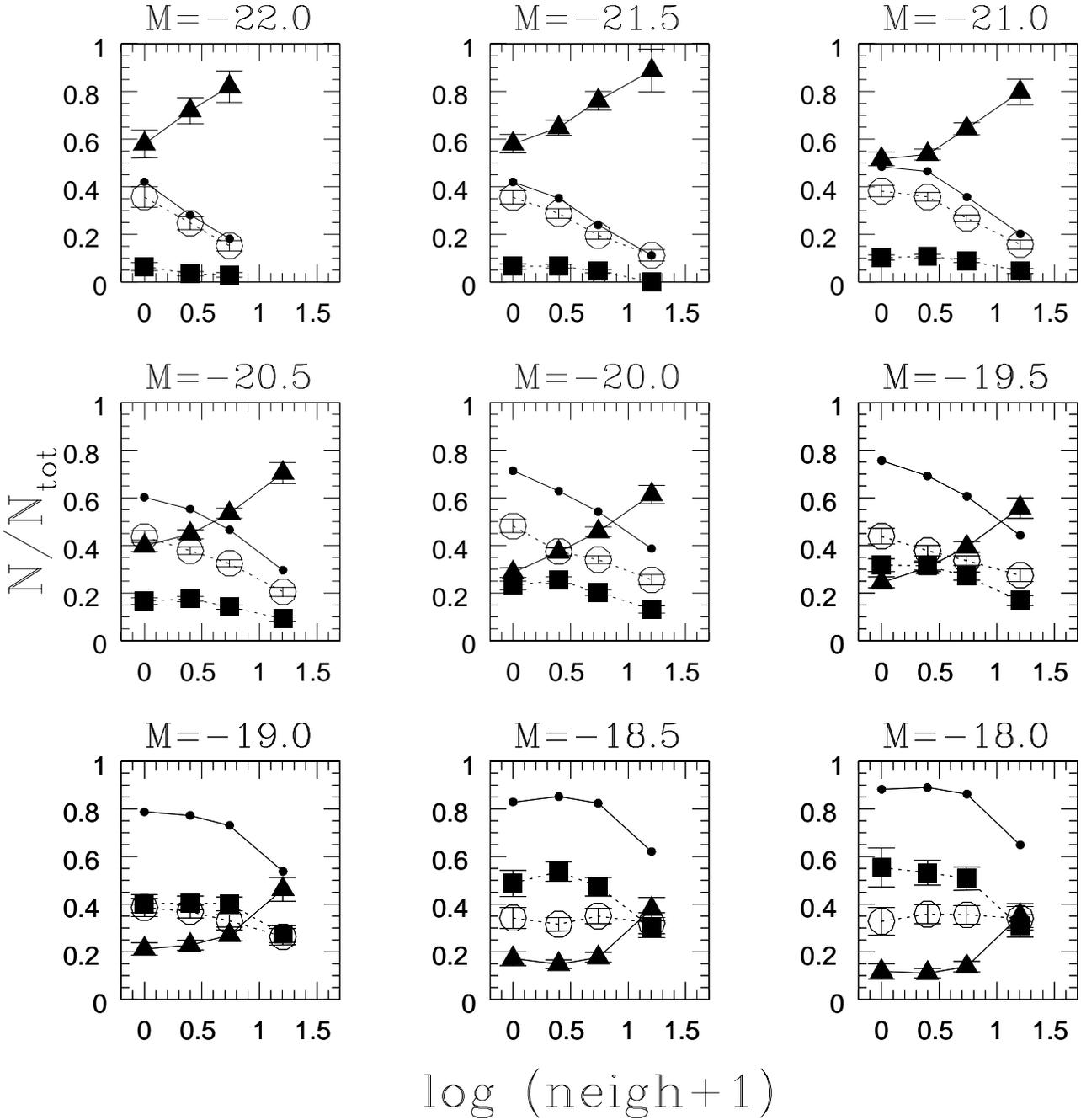}}
\hfill
\caption[]{The spectral-type/satellite-density relation for 
passive, quiet-SF and active-SF galaxies in the 9 brightest samples. 
Symbols are as in Fig. 2, small dots indicate the 
cumulative contribution of active and quiet-SF galaxies. 
Fractions have been computed using values listed in Table 2,   
normalized to the total number of galaxies in a 
given density bin so that, in each environment, the sum over 
all spectral types is 1.  
}
\end{figure*} 
Figures 5 provides evidence that 
the fractional increase of passive galaxies between extreme densities is a 
universal characteristic of galaxies that does not depend on luminosity. 
What depends on luminosity is the fractional increase (decrease) of passive 
(SF) galaxies between neigh=0 and a group-like density (neigh=3-8). 
The increase (decrease) is not observed in samples fainter than $\simeq$ --19. 
This supports our claim that the spectral-type/density relation is actually a 
process linked to the accretion of satellites by large massive halos, 
and not to enhanced correlation between distinct halos. 

Figure 4 and 5 also indicate that the efficient formation of faint passive 
galaxies is a threshold process that only occurs in systems where a galaxy 
has a very large number of neighbours; it does not proceed gradually with 
neighbour density. 
Clearly this suggest that mechanisms acting only in clusters (ram pressure and 
stripping) are more efficient than those acting also in groups (galaxy 
interactions) in generating faint passive galaxies.  

Finally Fig. 5 indicates that the neighbour density range corresponding to the 
intersection between early-type and late-type galaxies moves towards richer 
regions as the luminosity of the samples decreases.   
At M$_{B}$\,- 5 $\log$ $h_{75}$ =  -22 passive galaxies 
appear more numerous than SF galaxies even at the lowest densities (neigh=0).  
At M$_{B}$\,- 5 $\log$ $h_{75}$ = -20  equipartition is 
reached at group-like densities (neigh=3-8). At 
M$_{B}$ \,- 5 $\log$ $h_{75}$ = -18 equipartition  is never reached; passive 
galaxies are no more than one-third of the population even in the densest 
environment (neigh$>$8). 
This confirms that, on the group scale, luminosity generally dominates over 
environment in setting the spectral type mix of a population.  
The result is complementary to the finding (\cite{Norberg02a}) 
that luminosity, and not type, is the dominant factor in determining how 
the clustering strength of the whole galaxy population varies with 
luminosity. However, because in luminous (faint) samples galaxies are 
mainly passive (active-SF), the correlation with type is strong at both the 
high and low luminosity end. 
  
Given the correlation between spectral-type and morphology (\cite{Madgwick02}),
Fig.5 also likely illustrates a strong dependence of the morphology/density 
relation (\cite{Davis76,Dressler80,Postman84,Maia90,Helsdon03}) on luminosity. 
Nevertheless, to prove this dependence for the morphology/density relation
might be difficult as the correlation between environment and stellar age 
(color and spectral-type) appears stronger than the correlation 
between environment and morphology (\cite{Willmer98,Kauffmann04,Blanton05a}). 
\section{Summary and conclusions}
In this paper we have investigated relations linking the spectral-type 
properties of 2dF galaxies to their luminosity and local neighbour density 
characteristics. 
We have assigned a local density to galaxies computing the number of 
neighbours within a 1 $h_{75}^{-1}$ Mpc projected distance 
and $\pm$1000\,km\,s$^{-1}$ depth. Our approach differs from previous analyses dealing with the  
same issue, because we have computed densities counting only fainter neighbours 
and applying a maximum 
magnitude difference criterion (--2 $\leq$ M$_{gal}$ - M$_{neigh}$ $\leq$0). 
This implies that, at least for luminous
galaxies, neighbours  likely trace the density of satellites that have 
been captured by the galaxy  halo. 

We have shown that the local density distribution for the all-type 
galaxy sample is approximately luminosity-independent over the whole 
explored luminosity range. This indicates that, at least on a  1$h_{75}^{-1}$ Mpc scale, 
the number of neighbours associated with a galaxy is very similar, whether it reflects 
the number of satellites accreted by a luminous galaxy halo or  
the number of neighbours of low luminosity galaxies still in their original 
small halos. 

We have also found that that the excess of fainter companions 
surrounding passive galaxies is not limited to luminous galaxies 
(\cite{Norberg02a,Hogg03,Berlind04,Blanton05a})
but is instead a general characteristic of passive galaxies. 
The stronger clustering of passive galaxies relative to SF galaxies, 
on the group scale, appears to arise from two distinct contributions. 
At the luminous end, it is due to an excess of satellites surrounding 
central galaxies inside large halos. At the faint end it is due to an excess 
of satellites that are strongly correlated among themselves.  

We have shown that the global spectral-type/luminosity relation (Fig.2)
is not significantly altered in 
subsamples exhibiting different satellite densities. 
Whatever the environment, passive galaxies (Type 1)  
numerically dominate in luminous samples, and active-SF galaxies 
(Type 3+4) numerically dominate in faint ones. 
In contrast, the relative content in quiet-SF galaxies shows a 
weak dependence on luminosity. 
Only galaxies in extreme environments exhibit significant departures from 
these general trends: in dense environments (neigh$>$8)  a significant 
fraction of passive galaxies is observed even among faint galaxies, 
whereas among isolated galaxies (neigh=0), quiet-SF (Type 2) galaxies still 
represent a 40\% fraction of the luminous population. 
We suggest that these   
''minority'' populations, identified in very luminous and very faint samples, 
are poor tracers of halo mass: luminous SF galaxies are actually tracers of 
small  (sub-group scale) halos, whereas faint passive galaxies are tracer of  
massive (group/cluster scale) halos. 

Our analysis provides evidence for the existence of a global 
spectral-type/satellite-density relation, with the fraction of 
passive galaxies steadily growing (and the fraction of quiet-SF galaxies 
steadily decreasing) when moving from an isolated galaxy sample to galaxies 
with cluster-like neighbour density. 
But we have also shown that this relation only holds in luminous samples; 
in faint samples    
the variation in the fractional content of passive (SF) between the 
neigh=0 and the intermediate dense (neigh=3-8) environments is not observed:  
the dependence on environment becomes threshold-like, and very dense 
environments are required to observe a variation in the spectral-type mix. 
This suggests that the morphology/density relation is likely 
a morphology/satellite-density relation, that traces enhanced correlation 
inside single massive halos rather than enhanced correlation between 
distinct halos.  
\begin{acknowledgements}
We thank  A. Berlind, A. Biviano, R. De Propris, T. Goto  and  C.N.A. Willmer
for comments and suggestions. We are also indebted to the anonymous
referee whose comments and criticism greatly improved the scientific content of this paper. This
work was supported by MIUR, BK acknowledges a fellowship from Bologna University.
\end{acknowledgements}
{}

\begin{thebibliography}{}

\bibitem[Balogh et al. 2004a]{Balogh04a} 
Balogh, L.M., Eke, V., Miller, C.J. et al. 2004a, MNRAS, 348, 1355  

\bibitem[Balogh et al. 2004b]{Balogh04b} 
Balogh, L.M., Baldry, I.K., Nichol, R. et al. 2004b, ApJ, 615, 101 

\bibitem[Berlind et al. 2004]{Berlind04}
Berlind, A.A., Blanton, M.R., Hogg, D.W. et al. 2004 [astro-ph/0406633]
  
\bibitem[Blanton et al. 2003b]{Blanton03b}
Blanton, M.R., Hogg, D.W., Bahcall, N.A. et al. 2003, ApJ 594 186

\bibitem[Blanton et al. 2005a]{Blanton05a}
Blanton, M.R., Eisenstein, D.J., Hogg, D.W. et al. 2005a, ApJ, in press [astro-ph/0310453]

\bibitem[Blanton et al. 2005b]{Blanton05b}
Blanton, M.R., Eisenstein, D.E., Hogg, D.W., \& Zehavi, I. 2005b, ApJ, subm. 
[astro-ph/0411037] 

\bibitem[Christlein \& Zabludoff 2005]{Christlein05}
Christlein, D., \& Zabludoff, A.I. 2005 [astro-ph/0411359]

\bibitem[Colless et al. 2001]{Colless01}
Colless, M.M., and the 2dFGRS team 2001, MNRAS, 328, 1039 

\bibitem[Colless et al. 2003]{Colless03}
Colless, M.M., and the 2dFGRS team 2003, [astro-ph/0306581] 
 
\bibitem[Davis \& Geller 1976]{Davis76}
Davis, M., Geller, M.J. 1976, ApJ, 208, 13

\bibitem[Dominguez et al. 2002]{Dominguez02} 
Dominguez, M.J., Zandivarez, A.A., Martinez, H.J., et al. 2002, MNRAS, 335, 825

\bibitem[Dressler 1980]{Dressler80} 
Dressler, A. 1980, ApJ, 236, 351

\bibitem [Eke et al. 2004a]{Eke04a}
Eke,V.R, Baugh C.M., Cole, S., and the 2dFGRS team 2004a, MNRAS, 348, 866
 
\bibitem [Eke et al. 2004b]{Eke04b}
Eke,V.R, Frenk, C.S., Baugh, C.M., and the 2dFGRS team 2004b, MNRAS, 348, 1355 
    
\bibitem[Gerken et al. 2004]{Gerken04}
Gerken, B., Ziegler B.L., Balogh, M. L., et al. 2004 A\&A  421, 59

\bibitem[Gomez et al. 2003]{Gomez03}
Gomez, P.L., Nichol, R.C., Miller, C. J. et al. 2003, ApJ, 584, 210

\bibitem[Goto et al. 2002]{Goto02}
Goto, T., Hokamura, S., McKay, T., et al. 2002, PASJ, 54 , 515  

\bibitem[Goto et al. 2003]{Goto03}
Goto, T., Yamauchi, C., Fujita, Y., et al. 2003, MNRAS, 346, 601 

\bibitem[Helsdon \& Ponman 2003]{Helsdon03}
Helsdon, S.F., \& Ponman, T.J. 2003, MNRAS, 339, L29

\bibitem[Hogg et al. 2002]{Hogg02}
Hogg, D.W., Blanton, M., Strateva, I. et al. 2002, AJ, 124, 646

\bibitem[Hogg et al. 2003]{Hogg03}
Hogg, D.W., Blanton, M.R., Eisenstein, D.J. et al. 2003, ApJ, 585, L5

\bibitem[Hubble \& Humason 1931]{Hubble31}
Hubble, E., \& Humason, M.L. 1931, ApJ, 74, 43 

\bibitem[Jing \& Borner 2004]{Jing04}
Jing, Y.P., \&  Borner, G. 2004, ApJ, 617, 782 

\bibitem[Kauffmann et al. 2004]{Kauffmann04}
Kauffmann, G., White, S.D.M., Heckman, T.M., et al. 2004, MNRAS, 353, 713 

\bibitem[Kelm \& Focardi 2004a]{Kelm04a} 
Kelm, B., \& Focardi, P. 2004a, A\&A, 418, 937

\bibitem[Kelm \& Focardi 2004b]{Kelm04b} 
Kelm, B, \& Focardi, P. 2004b, in IAU coll. 195, Outskirts of galaxy clusters: 
intense life in the suburbs, ed. A. Diaferio, 456

\bibitem[Lewis et al. 2002]{Lewis02}
Lewis, I., Balogh, M., De Propris, R. et al. 2002, MNRAS, 334, 673

\bibitem[Maddox et al. 1990]{Maddox90} 
Maddox, S.J., Efstathiou, G., \& Sutherland W.J. 1990, MNRAS, 246, 433

\bibitem[Madgwick et al. 2002]{Madgwick02}
Madgwick, D.S., Lahav, O. Baldry, I.K. et al. 2002, MNRAS, 333, 133

\bibitem[Madgwick et al. 2003]{Madgwick03}
Madgwick, D.S., Hawkins, E., Lahav, O. et al. 2003, MNRAS, 344, 847

\bibitem [Maia \& da Costa 1990]{Maia90}
Maia, M.A.G., \& da Costa, L.N. 1990, ApJ, 352, 457   

\bibitem[Mateus \& Sodr\'e]{Mateus04}
Mateus, A.J., \& Sodr\'e, L.J. 2004, MNRAS, 349, 1251

\bibitem[Merch\'an \& Zandivarez 2002]{Merchan02}
Merch\'an, M., \& Zandivarez, A. 2002 MNRAS 335, 216

\bibitem[Norberg et al. 2002a]{Norberg02a}
Norberg, P., Baugh, C.M., Hawkins, E. et al. 2002a, MNRAS, 332, 827 

\bibitem[Norberg et al. 2002b]{Norberg02b}
Norberg, P., Baugh, C.M., Hawkins, E. et al. 2002b, MNRAS, 336, 907 

\bibitem[Mulchaey et al. 2003]{Mulchaey03} 
Mulchaey, J.S., Davis, D.S., Mushotzky, R.F., \& Burstein, D. 
2003, ApJS, 145, 39

\bibitem[Postman \& Geller 1984]{Postman84}
Postman, M., \& Geller, M.J. 1984, ApJ, 281, 95

\bibitem[Strateva et al. 2001]{Strateva01}
Strateva, I., Zeljko, I., Knapp, G.R. et al. 2001, AJ 122, 1861 

\bibitem[Tanaka et al. 2004]{Tanaka04}
Tanaka M., Goto T., Sadanori O., Shimasaku K., \& Brinkmann J. 2004, ApJ, 128,
2677

\bibitem[Tran et al. 2001]{Tran01}
Tran, K.H., Simard, L., Zabludoff, A.I., \& Mulchaey J.S. 2001, ApJ 549, 172 

\bibitem[Whitmore \& Gilmore 1991]{Whitmore91}
Whitmore , B.C., \& Gilmore, D.M. 1991, ApJ, 367, 64

\bibitem[Whitmore et al. 1993]{Whitmore93}
Whitmore , B.C., Gilmore, D.M., \& Jones, C. 1993, ApJ, 407, 489

\bibitem[Whitmore 1995]{Whitmore95}
Whitmore , B.C. 1995, in ASP Conf. Ser. 70, Groups of Galaxies, ed. 
O.G. Richter \& K. Borne, 41 
 
\bibitem[Willmer et al. 1998]{Willmer98}
Willmer, C.N.A., da Costa, L.N., \& Pellegrini P.S. 1998, AJ 115, 869 

\bibitem[Zehavi et al. 2002]{Zehavi02}

Zehavi, I., Blanton, M.R., Frieman J.A., et al. 2002, ApJ, 571, 172   

\bibitem[Zehavi et al. 2004]{Zehavi04}
Zehavi, I., Zheng Z., Weinberg D.H., et al. 2004, [astro-ph/0408569]   
 
\end{thebibliography}
\end{document}